# Photoemission Spectroscopy on photoresist materials: A protocol for analysis of radiation sensitive materials

Faegheh S. Sajjadian[1,2,a)], Laura Galleni[1,2], Kevin M. Dorney[2], Dhirendra P. Singh[2], Fabian Holzmeier[2], Michiel J. van Setten[2], Stefan De Gendt[1,2], and Thierry Conard[2,a)]

[1]Department of Chemistry, KU Leuven, Celestijnenlaan 200F, 3001 Leuven, Belgium
[2]Imec, Kapeldreef 75, 3001 Leuven, Belgium

a) Electronic mail: Faegheh.Sajjadian@imec.be, Thierry.Conard@imec.be

Device architectures and dimensions are now at an unimaginable level not thought possible even 10 years ago. The continued downscaling, following the so-called Moore's law, has motivated the development and use of extreme ultraviolet (EUV) lithography scanners with specialized photoresists. Since the quality and precision of the transferred circuit pattern is determined by the EUV induced chemical changes in the photoresist, having a deep understanding of these chemical changes is of pivotal importance. For this purpose, several spectroscopic and material characterization techniques have already been employed so far. Among them, photoemission can be essential as it not only allows direct probing of chemical bonds in a quantitative way but also provides useful information regarding the generation and distribution of primary and secondary electrons. However, since high energy photons are being employed for characterization of a photosensitive material, modification of the sample during the measurement is possible and this must be considered when investigating the chemical changes in the photoresist before and after exposure to EUV light.

In this paper, we investigate the chemical changes occurring during the photoemission measurements of an unexposed, model Chemically Amplified Resist (CAR) based on the well-known Environmentally Stable Chemically Amplified Photoresist (ESCAP) as a function of a number of measurement parameters, using both an X-ray (AlKα, 1486.6 eV) and UV source (HeII, 40.8 eV). We will show that these chemical changes can be simulated through theoretical modelling of the photoemission spectra. Based on these results, we propose a measurement protocol allowing to minimize or eliminate this modification which will help to guide (or enable) photoemission measurements on radiation-sensitive materials (e.g., photoresists).

# I.  Introduction:

Continuous downscaling of smart electronic devices while enhancing their processing power, based on the so-called Moore's law, has been facilitated through nanolithography during the last 30 years. In this technique, complicated electronic circuit patterns are printed to a few tens of nanometers of dimension. Originally, Ultraviolet (UV, from 436 nm to 193 nm) or Deep Ultraviolet (DUV, from 254 to193 nm) light transfers through a mask (with the desired circuit pattern), a lens-based illumination system (to reduce the dimension) and impact a photosensitive material called a photoresist. Hence, the photoresist images the pattern of the circuit through photochemical changes induced by the UV light. A development procedure followed by subsequent etching of the exposed area helps to transfer this pattern to a substrate of dielectric semiconductive and/or conductive materials. The dimensions of the pattern, as Rayleigh criterion states, can be controlled by the wavelength of the light, the numerical aperture of the imaging system, and other parameters[1]. The growing interest in achieving smaller patterns has motivated the use of shorter wavelength light i.e., Extreme Ultraviolet (EUV,13.5nm, 92 eV) instead of UV. Moving from UV to EUV regime lithography triggers the demand for the development of specialized next generation EUV-photoresists.

One of the common types of photoresists in the UV/DUV regime is the Chemically Amplified Resist (CAR) which have been modified for the EUV regime as well. A CAR consists of a polymer backbone as the main building block, a photoacid generator (PAG) and a quencher. Traditionally, in DUV lithography, photoactivation of CAR usually starts with the activation of PAG with incident photons, that leads to release and diffusion of acid in the system. This acid reacts with the polymer through a deprotection process which results in the solubility change of the exposed area of the CAR[1]. In the EUV-regime, due to higher energy of photons (~92 eV) compared to UV light (4 to 7 eV), different activation mechanisms are expected. The incident photons might either directly react with the polymer chain -due to the highest molecular cross section in the system- and activate the deprotection process, or they might generate photoelectrons from polymer which will go on to activate the PAG and other molecules, causing side reactions that deviate from the ideal PAG-mediated deprotection pathway[1].

Also, optimization of the photoresist performance is conventionally done by balancing the tradeoff between the pattern resolution, line edge roughness, and the photoresist sensitivity (known as RLS tradeoff)[1]. For EUV-photoresists, this optimization is challenging due to the very small dimension of the patterns, low power of the EUV sources and high amount of energy carried by photons[1–3]. The latter also causes a different photochemistry than what is known for the UV-photoresists as explained above, which is still not well understood[1]. Therefore, developing a deep knowledge of the EUV induced chemistry in these resists is essential for improving the photoresist performance.

To this end, several characterization techniques have been employed so far to shed some light on different aspects of this complicated radiochemistry. In a work by Kostko, et al.[4], three different measurements techniques were used to investigate the EUV-material interaction for several model polymer systems. First, the efficiency of the material in absorbing the incident EUV photons was evaluated by EUV reflectivity measurements and calculation of the EUV absorption coefficient which is an indication of the sensitivity of the photoresist. In the second step, the interest was to identify how efficient the polymer material is in converting the absorbed photons to photoelectrons. This was studied by looking at the emitted electrons through photocurrent measurements and calculation of the total electron yield (TEY). In the last step, the electron attenuation length (EAL) was calculated by measuring the drain current as a function of time. The EAL was determined as an indication of the distance that slow electrons can travel inside the material before collision. This is an important factor because in the EUV-CAR, the chemistry is supposed to be induced by slow electrons. Since electrons move randomly in the CAR, the longer travel in the system, enhances the chance of initiating unwanted chemical changes in the non-exposed areas of the material which degrades the resolution of the image[1,4].

Among other widely used techniques, one can also consider Fourier Transform Infrared Reflection (FTIR)[5] spectroscopy, which through changes in the vibrational spectrum of the chemical bonds, can give insight into the chemical state of a photoresist system pre- and post-EUV exposure. In addition, residual gas analysis (RGA)[6,7], can provide indirect information about the occurring chemical reactions, through the analysis of escaping molecular fragments. In the evaluation of photochemistry, RGA especially helps by providing direct information about the gas residuals from PAG fragmentation and polymer deprotection, gives indirect information about the activation of PAG. Cyclic voltammetry (CV) is another widely used characterization technique for the study of the reduction potential of PAG in a liquid phase by measuring the current versus the voltage applied on the working electrode. It is shown that higher reduction potential of the PAG results in improved sensitivity of the photoresist[7–9]. Electrolysis could assist as a complementary technique to CV which can be used to study the interaction between photoresist components and low energy electrons, providing information about the activation of the PAG and acid generation[7].

Another useful technique is X-ray photoemission spectroscopy (XPS) during which high energy X-ray photons (AlKα 1486.6 eV) ionize electrons out of the core shell electronic orbitals. A unique aspect of this technique is allowing direct probing of chemical bonds of individual elements in a quantitative way by providing kinetic energy and abundance of the generated photoelectrons upon x-ray exposure[4]. This could provide a unique advantage in step-by-step tracking of a chemical reaction or retrieving the empirical formula of an unknown material. For instance, in a work by Li, et al., the structure of hafnium methacrylic acid (HfMAA) hybrid nanoparticles - a type of photoresists which does not belong to the CAR family- after exposure to UV light was investigated by XPS[10]. The results showed a decrease in the mass percentage of the carboxylic group with increasing UV exposure time.

However, care should be taken when using XPS for probing photosensitive materials (like photoresists) since high energy X-ray photons can introduce unwanted chemical modification during the measurement which has been studied and reported in several works[11–14]. In a work by Pan, et al., XPS was used to probe, step by step, the X-ray induced decomposition of perfluoropolyether (PFPE) lubricant applied to a gold film[15].

That being the case, modification of polymeric materials by XPS measurements is a great challenge for the investigation of Chemically Amplified Resists (CAR). Considering the energy of X-ray photons (which is far higher than the EUV photons), modification of the sample during the measurement is possible and this must be considered when investigating the chemical changes in the photoresist before and after exposure to EUV light.

As a complementary technique to XPS for investigation of photoresists, Ultraviolet-Photoemission Spectroscopy (UPS) can be mentioned which is typically performed with either HeI(20.2 eV) or HeII (40.8 eV). These lower energy photons (compared to X-ray), extract photoelectrons from the valence orbitals, where most chemical bonds form/decompose. Therefore, UPS can provide information about the valence band structure, the position of Fermi level and work function of the material. However, care should be taken when using UPS for investigation of CARs, since the above-mentioned explanation about the modification of photoresists during XPS measurements holds true for the UPS measurements as well[16].

In this research, we investigate the chemical changes occurring during the photoemission measurements using both an X-ray (AlKα 1486.6 eV) and UV source (HeII 40.8 eV), of an unexposed, model CAR based on the well-known ESCAP platform, as a function of measurement parameters including X-ray photon density, the use of a charge neutralization system, beam spot size and measurement duration. Atomic force microscopy (AFM) and XPS elemental mapping are complementarily employed to interpret the nature of the chemical modification detected after XPS measurement. We will show that the photochemistry induced by both XPS and UPS, can be understood through theoretical modelling of the photoemission spectra[17]. Based on these results, we propose a measurement protocol allowing to minimize or eliminate this modification. This measurement protocol will help to guide (or enable) photoemission measurements on light-sensitive materials (photoresists) that result in minimal perturbation of the material itself. Later this protocol will be helpful for the study of exposed ESCAP to DUV/EUV, by disentangling the chemical changes induced by the measurement from those that are directly resulting from different steps of the DUV/EUV lithography processes (i.e., exposure, pre/post exposure bake, etc.).

## II.  Experimental Setup and Methodology:

XPS measurements were conducted on a model CAR consisting of poly[(t-butyl methacrylate)-co-(p-hydroxystyrene)] copolymer (44.9 mol% PBMA, 41.5 mol% PHS), mixed with (4-Methylphenyl) diphenyl sulfonium nonaflate (9 mol%) PAG, and trioctylamine (3 mol%) quencher. Furthermore, traces of two residual solvents (1.2 mol% Propylene glycol methyl ether acetate (PGMEA), 0.4 mol% Propylene glycol methyl ether (PGME)) are present in the resist formula (Figure 1). As a comparison, XPS spectra were also acquired on polymer-only (52 mol% PBMA, 48 mol% PHS), i.e., without PAG and Quencher. One series of samples were prepared by spin coating on a 200 mm silicon wafer or on $5 \times 5$ cm$^2$ silicon coupons in a clean room environment followed by a post application bake at 90 °C for 60 s. Later, another series of samples were prepared by spin coating on $1 \times 1$ cm$^2$ silicon coupons which were coated by 150 nm gold layer, under the same conditions. The gold layer helped with reducing the charging on the samples, no substrate induced chemical effects are expected for this system. Film thicknesses between 30 and 50 nm were obtained according to the recipe provided by the resist supplier.

| Polymer (86.4 mol%) | | PAG (9 mol%) | Quencher (3 mol%) | Solvent (1.6 mol%) |
|---|---|---|---|---|
| PHS | PBMA | 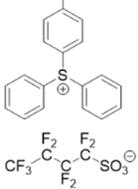 | 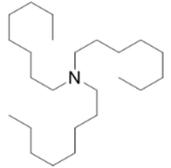 | 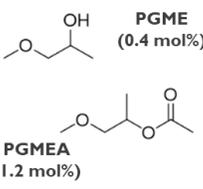 |
| 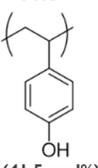 | 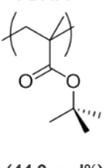 | | | PGME (0.4 mol%) |
| (41.5 mol%) | (44.9 mol%) | | | PGMEA (1.2 mol%) |

Figure 1: Model CAR used in this research. Beside the full CAR, samples of polymer-only were used.

The XPS and UPS measurements were performed employing a VersaProbe III and a Quantes hemispherical electron analyser from Ulvac-PHI using a monochromatized X-ray source (Al Kα, 1486.6 eV) and a He gas-discharge lamp (UV, He1 = 20.2 eV and HeII = 40.8 eV) photon beams. A flood gun providing low energy electrons (1 to 3 eV) was also used as a charge neutralization system whenever we have sample charging, to help the stabilization of the sample surface potential.

The samples were kept in the vacuum chamber prior to the measurements for 12 h to ensure that no significant outgassing in the vacuum would occur and that the vacuum level in the chamber could be maintained sufficiently low (in the range of $10^{-7}$ Pa) to carry out the photoemission measurements. The XPS and UPS measurements were performed at a take-off angle of 45° and 90°, respectively. The UV beam spot size is fixed and is about 1 mm$^2$.

To better evaluate the effect of X-ray induced damage on the samples, various photon doses were investigated. This was achieved both by varying the measurement time between 9 and 400 minutes and by using an X-ray spot size of 100×100 µm$^2$ either in a non-rastered mode or with rastering over an area of 1000×500 µm$^2$. The X-ray power was set to 25W. XPS elemental mapping was performed with the Quantes tool utilizing an X-ray beam of 20×20 µm$^2$ at 4.5W and a take-off angle of 45 deg on several damaged areas to detect traces of residual resist after XPS measurements.

Complementary to the XPS measurements, which reveal changes in the chemical state, the samples were also developed after exposure to emphasize possible traces of sample modification from XPS measurements. Here, a post exposure bake (PEB) at 90°C for 60s was applied, followed by development with tetramethylammonium hydroxide (TMAH- 2.62% aqueous solution) carried out for all samples. The exposed areas were then inspected under an Olympus DSX500 optical microscope which provides 10X magnification. The microscope is used in bright-field mode with HDR-enhanced image acquisition (HDR = High Dynamic Range) and utilizes the standard DSX image acquisition and processing software. Furthermore, AFM was employed using a r-ICON PT, Nanoscope V system to investigate changes in morphologies of selected samples.

To support the interpretation of XPS and UPS measurements, the theoretical spectra of the systems under study were simulated by means of first-principles calculations. For simplicity, the theoretical spectra for XPS and UPS were estimated by calculating the density of states (DOS) of the systems under study, assuming energy-independent photoionization cross-section and no energy-loss from scattering. The DOS was obtained by applying an arbitrary gaussian broadening of 0.6 eV to the calculated binding energies. All calculations were carried out using Turbomole 7.2[18].

The methodology to obtain the DOS has been reported in a previous work[17]. To reduce the computational cost, the DOS of the ESCAP material was approximated as a sum of the DOS of its separate molecular components, weighted by the molar ratios. Moreover, hydrogen-terminated monomers were considered instead of long copolymer chains. Our previous work showed that this approximation provides accurate results for XPS spectra of non-conjugated systems, thanks to the strong localization of core-electrons[17].

To calculate the binding energies, the atomistic structure of each component in the gas phase was first optimized in vacuum with density functional theory (DFT) using the PBE functional[19] and a Gaussian basis set of triple-$\zeta$ valence quality (def2-TZVP)[20]. Then, Kohn-Sham orbitals and energies were calculated with DFT at the BH-LYP/def2-SVP level of theory[20–23]. A single-shot GW calculation ($G_0W_0$) was then performed on top of the DFT calculation to estimate the binding energies[24].

The binding energies calculated with this protocol are referred to the vacuum level. Therefore, a rigid shift is needed to align the theoretical spectra with experiments, to compensate for shifting factors such as the work function, applied bias, charging effects, and residual discrepancies arising from approximations in the model. All theoretical C1s spectra were shifted to align the main C 1s peak at 284.8 eV. To simulate

UPS spectra, the same protocol has been used as for XPS. Theoretical binding energies were shifted to align the main peak around 8 eV in the experimental spectra.

# III. Results and Discussion:

## A.   XPS results

### 1.   XPS-induced damage of photoresist

First experiments were performed on a $3 \times 1$ cm$^2$ sample, cleaved from a 200 mm Si wafer coated with 35 nm of the ESCAP material. Short repeated XPS measurements of 20 minutes were performed, on three different positions, for three different total data collection durations of 60, 240 and 400 minutes. This allowed to evaluate the effect of long-time x-ray exposure on the sample in a comparable way. The measurement was carried out with an x-ray beam spot size of 100 µm diameter and no charge neutralization was used. Photoemission spectra were recorded for Carbon (C1s), Oxygen (O1s), Fluorine (F1s) and Sulphur (S2p) (Figure 2), which are the main elements present in the CAR. Nitrogen is another element present in the quencher component of the CAR, however recording the Nitrogen (N1s) spectra was not possible due to the very low concentration of this element in the sample. Recorded spectra were later calibrated by the conventional referencing method for the organic materials, using the main carbon peak at 284.8 eV, to compensate for any possible peak shift which might result from the surface charging during the measurement.

The first observation of the results is the pronounced decrease in the intensity of oxygen peak around 533 eV and fluorine at around 689 eV as well as the degradation of the minor carbon peak at 289 eV which is assigned to the carbonyl group. Also, for sulfur peak, with a more complicated situation, the decrease of the low binding energy (B.E.) component at 164 eV is accompanied by the increase of the high B.E component at 169 eV. On the other hand, the shape and broadness of the peaks also change when comparing the three measurements. The changes in the shape and intensity of the recorded spectra over time indicate sample modification during the XPS measurements.

Considering the chemistry of the model CAR as shown in Figure 1, any changes in XPS peaks of fluorine and sulfur implies PAG-involved photochemical changes, i.e., PAG degradation and acid generation. This would further result in the deprotection of the polymer (leading to the change in the solubility of the CAR material) which is expected to be visible in the spectra of C1s and O1s. However, since carbon and oxygen are present not only in polymer but in all components of the CAR, including PAG and quencher, it is not easy to directly distinguish the chemical changes from the spectra of these two elements. In addition, there is no guarantee that the polymer does not undergo direct chemical changes under X-ray exposure. This cannot be evaluated by only one XPS measurement on the CAR sample.

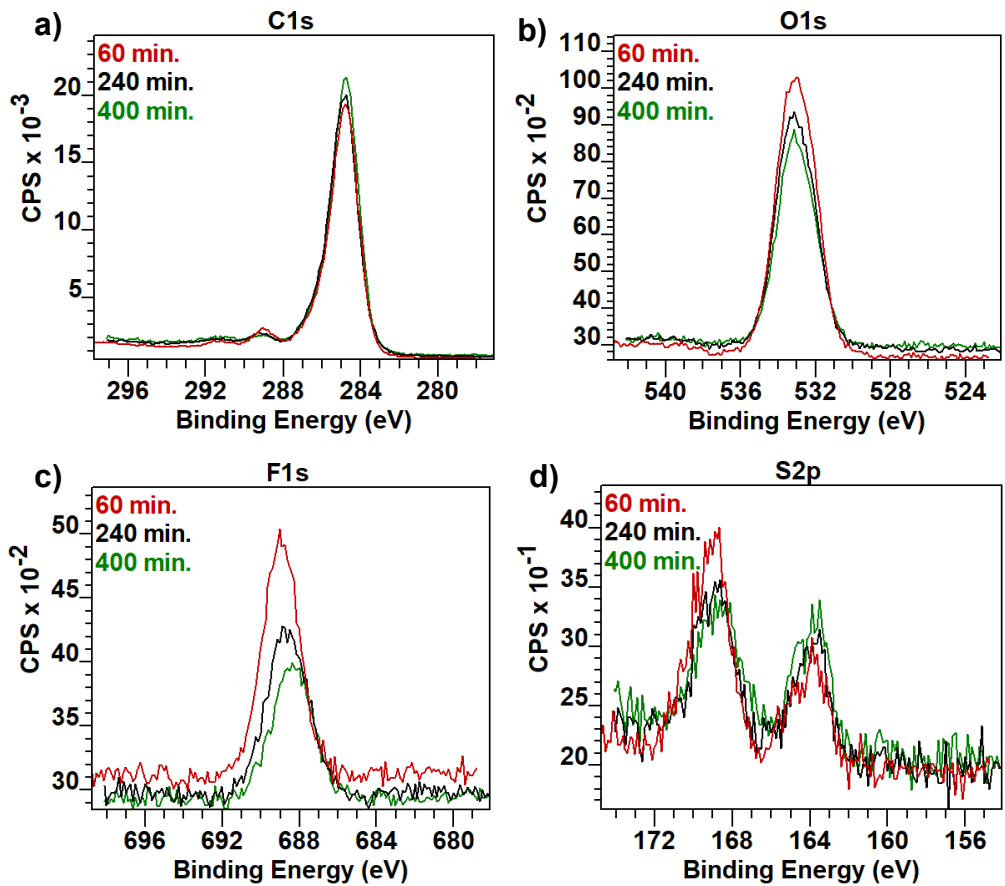

Figure 2: Short XPS spectra (recorded for 20 minutes) of the model resist measured after total exposure time of 60, 240 and 400 minutes for a) C1s; b) O1s; c) F1s and d) S2p.

In order to disentangle the effect on the polymer from the other components, another XPS measurement was performed on a polymer-only-sample of this model CAR (without PAG and Quencher). To keep consistency between the data for a valid comparison, the sample was measured in a similar experiment as for the full resist sample. A sample of $1 \times 1$ cm$^2$ coupon coated with 30 nm polymer, was measured with short repeated XPS measurements on three different positions for different total durations of 60, 240 and 400 minutes. Comparison between recorded spectra for C1s and O1s shows the degradation of the polymer-only sample under XPS measurement which is especially pronounced in terms of decrease in the C=O peak at 289 eV as shown in Figure 3a. This degradation is even more pronounced when we look at the corresponding relative atomic concentrations calculated for all the spectra recorded every 10 minutes during the total measurement time of 400 minutes as illustrated in figure 3c. This is in line with what was reported previously in another work, regarding the degradation of polymer-only sample by EUV exposure, which was revealed during an RGA measurement[25]. In the previous UV regimes of 248nm (5 eV) or 193nm (6.5 eV), the deprotection of the polymer happened through the reaction of the ester group with the generated acid from PAG excitation, thus for the polymer-only sample exposed to X-rays a totally different deprotection reaction should have happened. The suggested chemical reactions for this modification are direct scission of either the *tert*-butyl group which is the protecting group (PG) of the polymer chain, or the complete ester side chain, as illustrated in Figure 3d[25]. As shown in Figure 3a and c, the decrease of the C=O band at 289 eV in the C1s spectrum, and the relative decrease of the oxygen to carbon content of the system, might imply the scission of the whole side group.

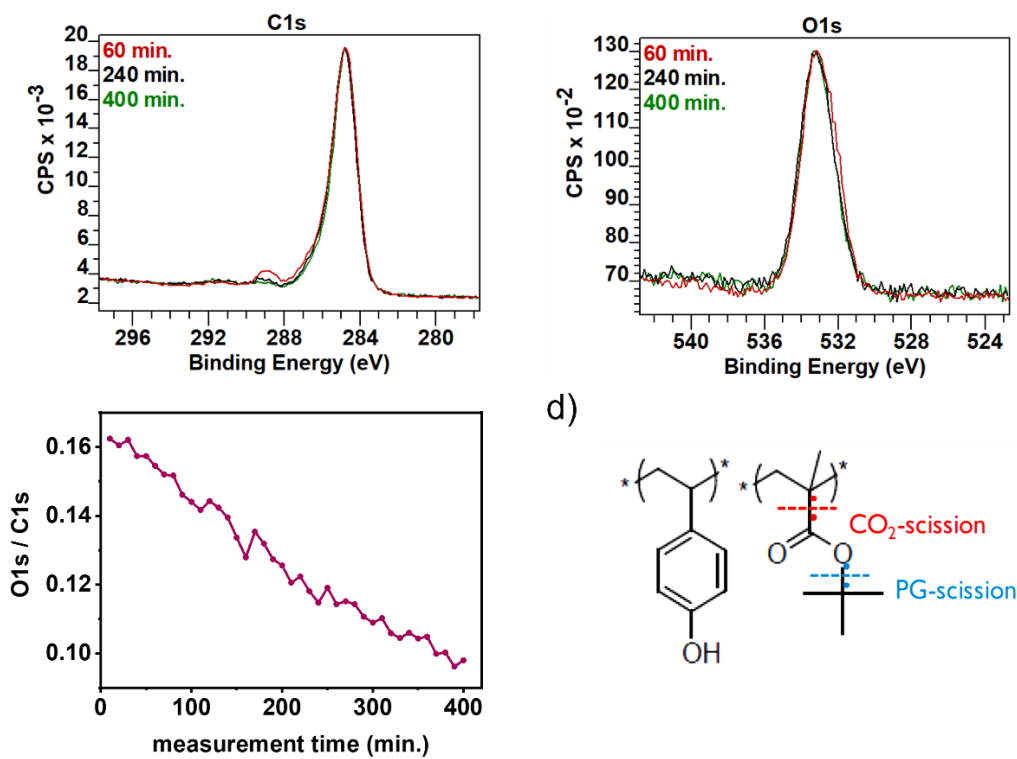

Figure 3: Short XPS spectra (recorded for 20 minutes) of the polymer-only sample measured after total exposure time of 60, 240 and 400 minutes: a) C1s spectrum and b) O1s spectrum c) corresponding relative atomic concentrations ratio for carbon and oxygen for all short spectra which are recorded every 10 minutes for total duration of 400 minutes; d) suggested activation reaction for under XPS measurement.

To gain insights into the chemical changes induced by XPS on model CAR and polymer-only sample, a theoretical simulation was carried out to predict the modification of C1s spectra upon sample degradation. Figure 4 and Figure 5 show the simulated XPS spectra of the full CAR and of the polymer-only sample, respectively, both before and after degradation. The degradation of the full CAR was modeled by assuming the photodissociation of the PAG into the two fragments shown in Figure 4c. For the polymer-only sample, it was assumed that one third of the PBMA monomers dissociate releasing $CO_2$.

As illustrated in Figure 4b, the simulated spectrum of the pristine ESCAP and the experimental spectrum after 20 minutes XPS measurement, shows a clear disagreement in the binding energy region between 286 to 290 eV, indicating that in the above-mentioned measurement condition, we are unable to record the pristine state of the model CAR because the sample modifies immediately after the start of the measurement.

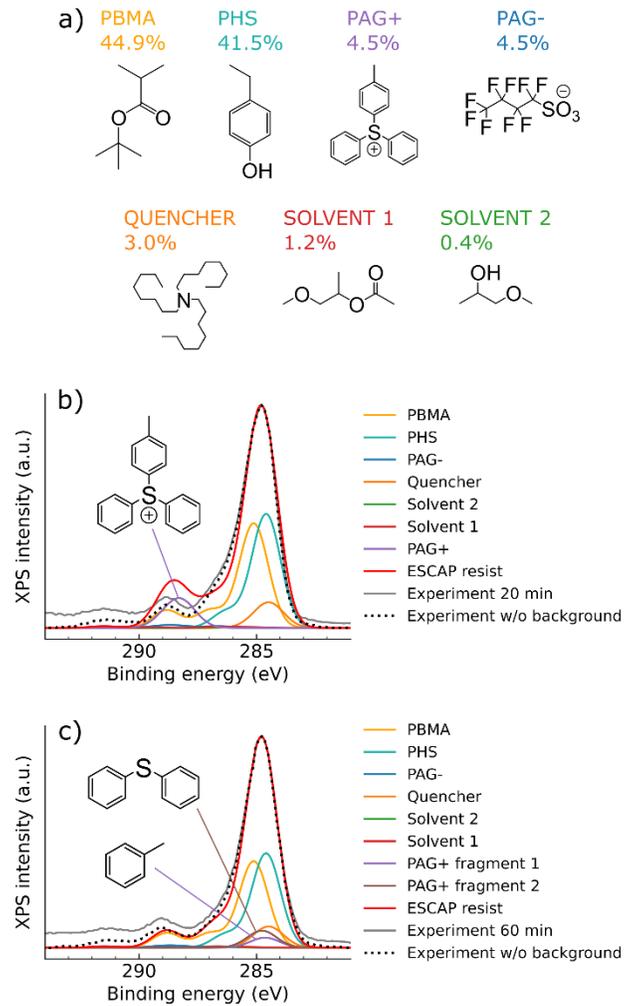

Figure 4: a) building blocks of the model CAR; b) Comparison between experimental C1s spectrum of the model CAR after 20 minutes and the theoretical model for pristine CAR; c) Comparison between experimental C1s spectrum of the model CAR after 60 minutes and theoretical model after PAG decomposition.

On the other hand, there is a good agreement between the simulated data for the full resist with decomposed PAG and the experimental spectrum recorded after 60 minutes XPS measurement. This agreement not only confirms the relatively fast degradation of the ESCAP material under XPS measurements, but also suggests that this degradation is due to the decomposition of PAG.

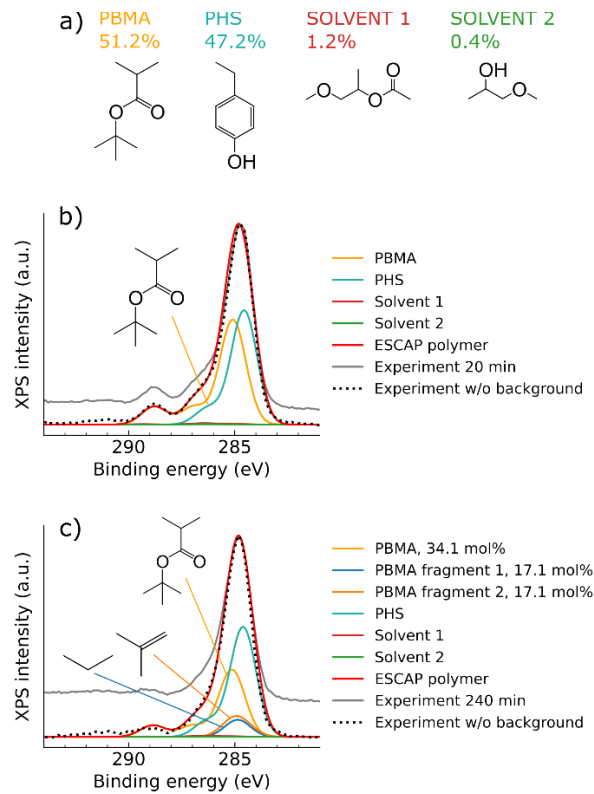

Figure 5: a) building blocks of the polymer-only sample; b) Comparison between experimental C1s spectrum of the polymer-only sample after 20 minutes and theoretical model; c) Comparison between Experimental C1s spectrum after 60 minutes and theoretical model after polymer fragmentation (dissociation of 33.3% of PBMA monomers).

For the polymer-only sample, we see a different situation in Figure 5. There is a good agreement between the simulated spectrum of the pristine polymer and the experimental spectrum recorded after 20 minutes XPS measurement in Figure 5b, which shows that the pristine state of the polymer-only sample is measurable with a short measurement in normal condition and confirms slower degradation of the polymer sample compared to the full resist.

On the other hand, as illustrated in Figure 5c, there is agreement between the simulated spectrum of the decomposed polymer and the experimental spectrum recorded after 240 minutes which confirms that there is far slower degradation for the polymer-only sample compared to the full ESCAP, under XPS measurements. This agreement also suggests a different degradation pathway than the one expected for the CAR sample, which is fragmentation of the PBMA component of the copolymer.

These results imply the demand for a precise measurement protocol to enable accurate and precise measurement, while preventing or limiting sample modification. On the other hand, since this photoresist is designed to be activated under UV/EUV radiation (photon energy range of below 100 eV), it is interesting to know how the sample reacts to the exposure by high energy X-ray photons (1486.6 eV).

To further understand the modification happening during XPS measurements, the measured samples of model CAR and polymer-only were submitted to a PEB and a development process and inspected further under the optical microscope to help with visualization of the traces of the chemical modification. It is noticeable that the ESCAP material used in this research is a positive tone resist and therefore, after the development process, the unexposed areas of the sample, appear in dark colour under the optical microscope which is an indication of the presence of the ESCAP. On the contrary, the light colour is an indication of the exposed area where the solubility of the ESCAP has changed after exposure to X-ray photons and PEB, and the model CAR is washed away during the development process. Keeping that in mind, the modification traces on model CAR, as depicted in Figure 6, are visible as surface damage with different areal sizes, which grow bigger with the measurement time. One common characteristic between all three detected spots, is that the damage area is not homogeneous and is recognizable as a light grey ring with a middle grey area in the centre, which is placed in the dark grey background of the non-exposed photoresist. Moreover, despite the X-ray beam spot size is ~100 µm diameter, the size of the damage spot is ~200 µm diameter for 60 minutes measurement and becomes larger for longer measurement durations, being ~300 µm diameter after 400 minutes XPS measurement. This larger damage spot than the beam spot, might indicate the spread of the chemical reaction to a larger area during the measurement. In the case of model CAR, one hypothesis would be that long X-ray exposure provides enough photon dose to activate regions of the resist that are outside of the main beam spot. Also this high photon dose could cause crosslinking of the deprotected polymer chains to a higher number of the neighbouring molecules resulting in growth in the size of the middle-dark central spot in the model CAR as well as the polymer-only sample[25]. However, for full CAR, the delamination of the sample by high energy x-ray photons is another possibility. This question remains open for further investigation.

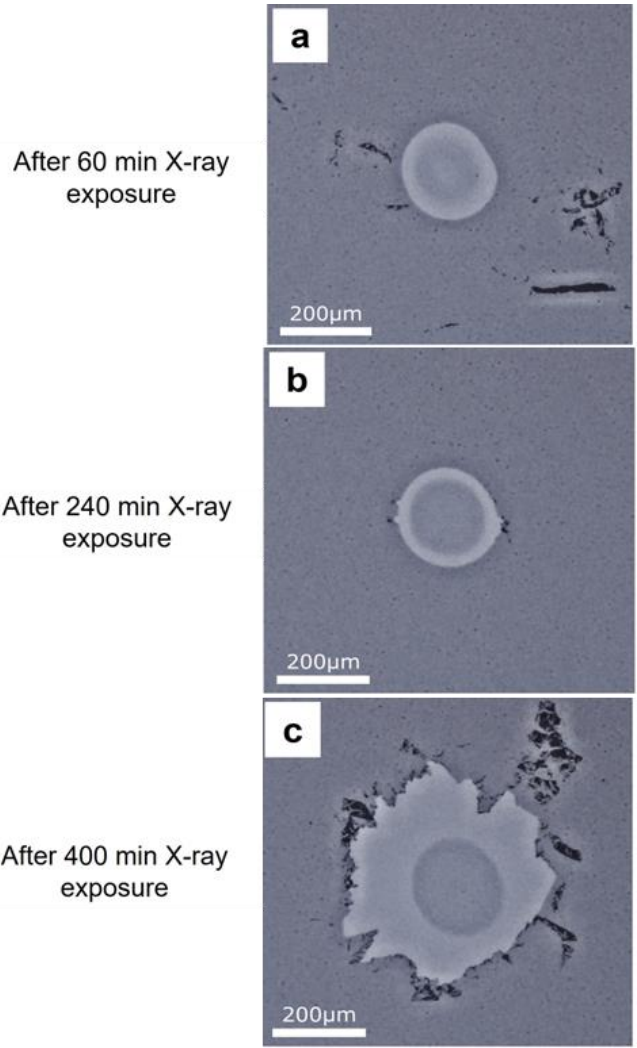

Figure 6: Inspection of the CAR sample under optical microscope after XPS measurement, PEB and development for different measurement durations: a) 60 minutes; b) 240 minutes; c) 400 minutes

On the contrary, for the polymer sample, as depicted in Figure 7, the damage spots do not grow much by increasing the total X-ray exposure duration, being ~205 µm after 60 minutes and growing to ~238 µm after 400 minutes measurement. Also, despite the similar shape of the damage spot recognizable as a light ring, with a darker colour in the centre, the size ratio of the light ring and the dark centre area is almost constant for different measurement times. These non-growing damage spots together with the chemical changes observed in the recorded spectra, might confirm the alternative direct scission reaction suggested for polymer deprotection in the absence of PAG and indicate that this reaction does not spread to far distances from the exposure spot. It is worth mentioning that the gold colour of the polymer samples is due to the gold layer coated on Si substrate.

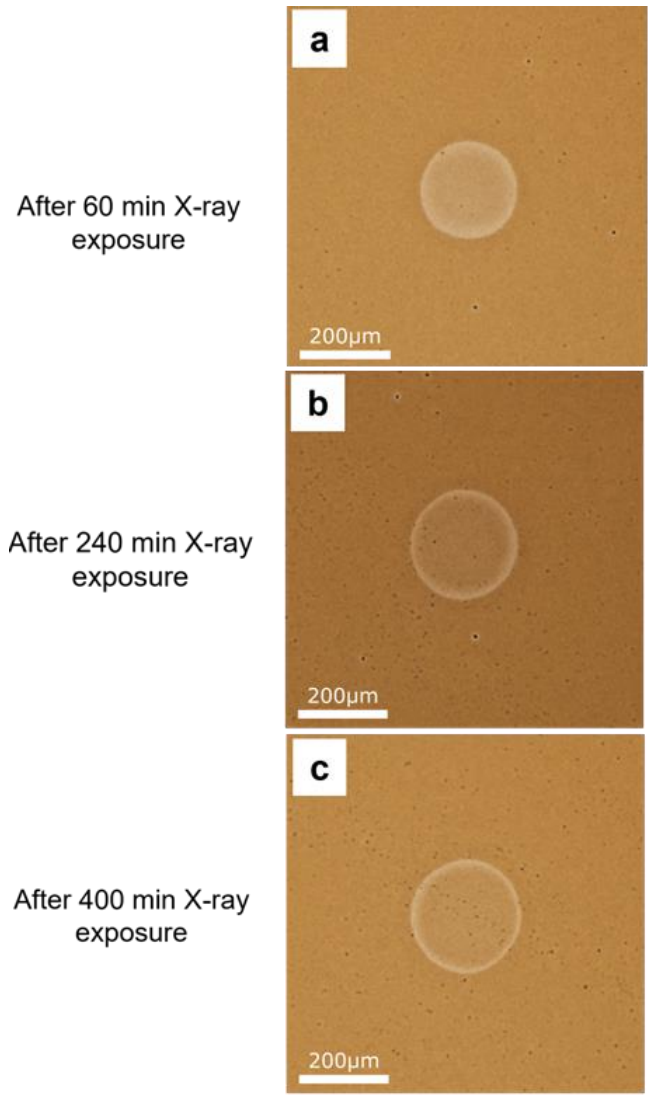

After 60 min X-ray
exposure

After 240 min X-ray
exposure

After 400 min X-ray
exposure

Figure 7: Inspection of Polymer-only sample under optical microscope after XPS measurement, PEB and development for different measurement durations: a) 60 minutes; b) 240 minutes; c) 400 minutes

In order to understand the nature of the light and middle grey colour areas in the model CAR, the damage spot after 60 minutes XPS measurements, followed by PEB and development, was inspected with AFM to obtain information about the morphology of the area and with XPS-mapping to retrieve corresponding elemental information with spatial resolution, as can be seen in Figure 8 and Figure 9, respectively.

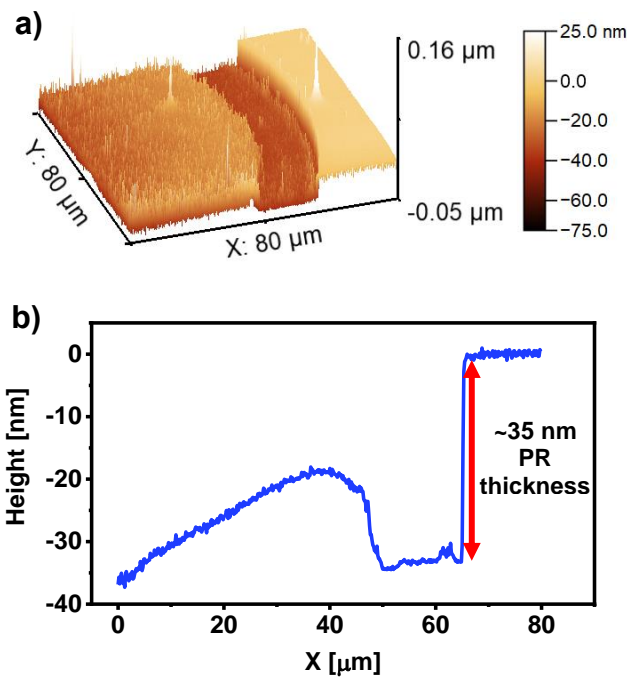

Figure 8: Damage spot on model CAR after 60 minutes XPS, development and PEB: a) 3D AFM image from a selected area of 80×80 µm² on the right edge of the modified area, b) corresponding average height of the area in a.

In Figure 8a, which shows a 3D AFM image from a selected area of 80×80 µm² on the right edge of the damage spot, the area of X-values between 60 to 80 µm in light brown indicates the highest altitude which is a sign of presence of non-exposed photoresist. Moving towards left, one can recognize the lowest altitude in dark brown, extended from x=60 to x=40 µm, with ~35 nm difference in height with respect to the light brown area as shown in the graph in Figure 8b. This dark brown region corresponds to the light grey ring in the optical microscope image shown in Figure 6a. This region might be an indication of the development and full removal of the photoresist resulting in the detection of the Si substrate. This theory is further confirmed by the XPS elemental mapping provided in Figure 9a, which clearly shows the detection of high concentration of silicon in this area.

Going further towards the centre of the damage spot in the above 3D AFM image in the x area between 40 to 0 µm, the altitude of the middle brown colour is between the values of the non-exposed resist area and the fully developed ring area. Also, the XPS mapping results for C1s at the centre of the damage spot, shown in Figure 9b, shows the presence of high amount of carbon which is more than contamination carbon and is most probably from the polymer backbone. These results together, indicate the presence of residual photoresist. This feature has analogy to the over-exposed resist to EUV light which results in residual cross-linked polymer[25]. However further investigation via other chemical characterization techniques or theoretical simulation is needed to better understand if the central area is partially developed resist or might be an indication of over-exposed cross-linked polymer due to exposure to high energy x-ray photons.

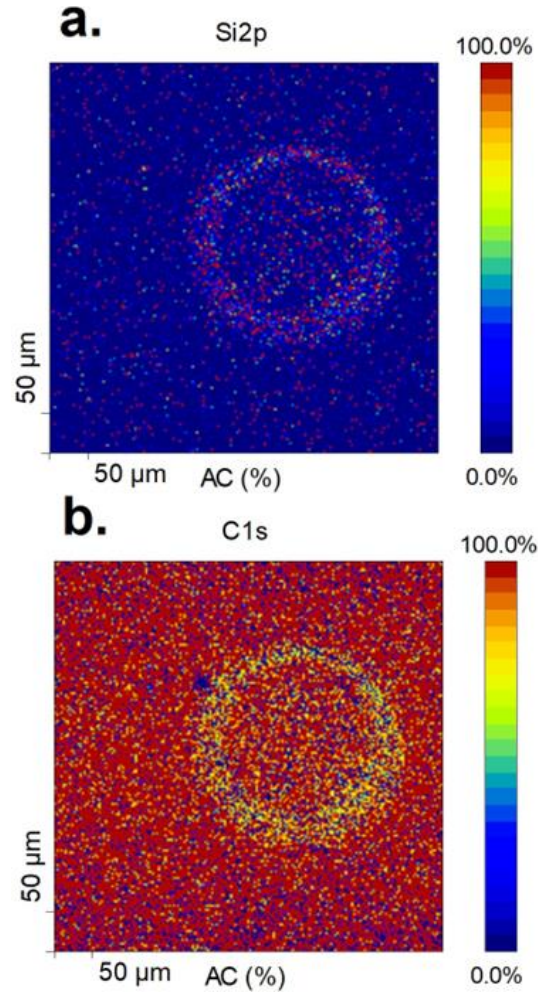

Figure 9: XPS elemental mapping on the damage area after XPS measurement for 60 minutes on ESCAP material for a) Si2p and b) C1s

## 2.    *Protocol for radiation damage-free XPS measurements of photoresist*

From the above given description, XPS-induced damage is linked to the photon exposure dose. It is thus needed to optimize our experimental procedure further to develop a measurement protocol to avoid modification during the measurement. This also implies that changing of the measurement parameters should be done while keeping the signal to noise ratio (SNR) of the recorded spectra at an acceptable level, especially for low concentration elements like fluorine and sulfur that could provide important information regarding activation of the PAG. As shown in Figure 6 the size of the damage in the model CAR, increases by increasing the measurement duration. Therefore, as a first step the effect of shorter measurement time in decreasing the sample damage during the measurement, was examined with the same beam spot size of 100 µm.

For this examination, XPS measurements conducted for different durations from 60 minutes down to 9 minutes on different positions of a sample of 2×2 cm$^2$ of the model resist by recording the spectra for C1s, O1s, F1s and S2p. Subsequent PEB and development carried out on the sample to help for further inspection of the exposure spots under the optical microscope and by XPS elemental mapping for C1s and Si2p. The results regarding the shortest measurement time of 9 minutes are shown in Figure 10. Corresponding S2p spectrum with poor SNR is a hint for not trying shorter measurement time. It is noticeable that although the XPS mapping does not show any traces of measurement damage, a pronounced spot is clearly visible in the optical microscope image, after development. This implies that while the exposure time to x-ray photons on the sample is reduced, still the measurement position is receiving a too high photon dose leading to degradation. That being the case, the second step is to try to reduce the amount of received photons on each point of the measurement position.

**9 minutes**

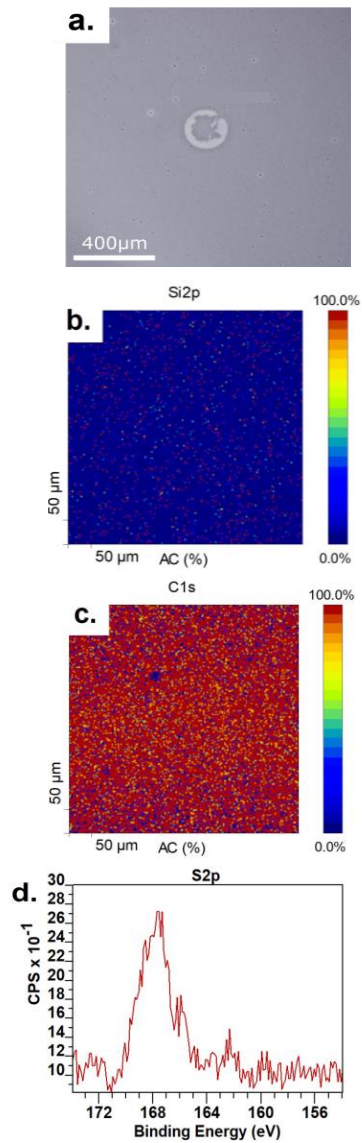

Figure 10: Model CAR after 9 minutes XPS measurement, PEB and development; a) Trace of sample modification under optical microscope; b) XPS elemental mapping for S2p; c) XPS elemental mapping for C1s; d) Recorded S2p XPS spectrum.

In our XPS setup, the photon flux as the number of photons emitted per second is fixed and hardly changeable. Thus, a substitute solution could be to spread the emitted photon flux from the source on a larger measurement spot. In this way the number of photons received per second per unit of surface is reduced. In our setup, this could be done by choosing a larger measurement spot size or preferably by raster scanning the beam during the measurement. We selected a measurement spot of $1000{\times}500\ \mu m^2$ was used which is ~50 times larger in area than the usual $100\ \mu m$ spot, causing a significant drop in the number of photons received per second per unit of surface.

Subsequently XPS measurements in those conditions were carried out for time varying between 45 minutes and 115 minutes. The sample was then submitted to the standard PEB and development process. The corresponding recorded S2p spectra and images under the optical microscope, and the results of XPS mapping are shown in Figure 11. While the results for 115 minutes measurement show obvious trace of measurement damage on the sample, for 45 minutes measurement, neither under optical microscope nor with the XPS mapping, no damage traces found, indicating this might be the optimal measurement condition from the damage point of view. However poor SNR for S2p recorded during this measurement, together with the extra broadening of the peak from both measurements, might be an indication of charging effect on the sample which might be improved by changing the sample mounting scheme.

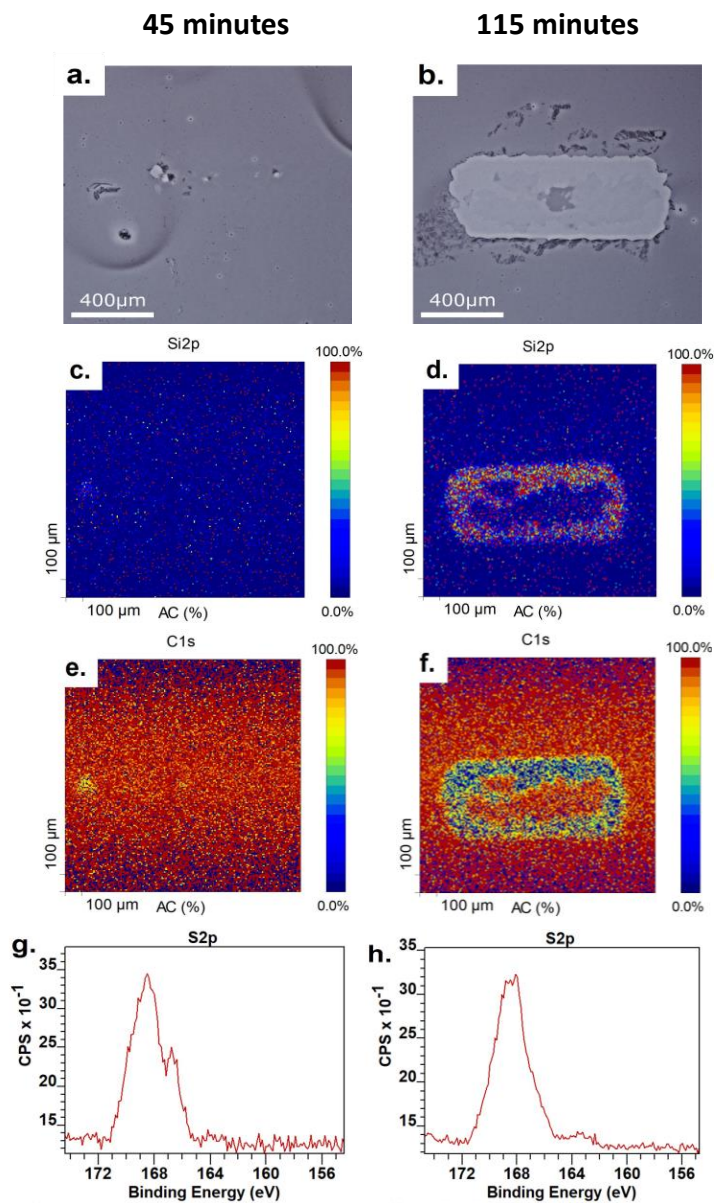

Figure 11: XPS on the model CAR with large X-ray spot size. Comparison between measurement times of 45 and 115 minutes: a and b) inspection under optical microscope after PEB and development; c and d) XPS mapping for Si2p; e and f) XPS mapping for C1s; g and h) The S2p spectra recorded to check the SNR.

## B. UPS results:

The study of valence band might be critical in understanding chemical modification in resists as most of the chemical bonds form or decompose. As the photoemission cross section in the valence band by XPS is low and that X-ray induced damage are easily present, the use of UPS might be of a greater help. UPS provides also information regarding the position of Fermi level and the work function of the material.

UPS measurements were carried out on Versaprobe with a measurement spot size of ~1 mm, at a take-off angle of 90 degree. First measurements were performed using HeI (20.21 eV) and the photon flux was limited by using a small aperture in the photon beam in order to limit possible charging effects. Unfortunately, no stable spectra could be obtained. This forces us to use electron charge neutralization in order to get stable spectra. However, low energy electrons from the neutralization system would easily damage the detection system, which forbid us to use electron neutralization in combination with HeI radiation. It also makes the determination of the work function impossible.

An alternative is to use HeII in combination with electron neutralization as in this case, the low energy electrons do not influence significantly the valence band measurements. Repeatability tests were conducted for HeII radiation with or without neutralization and with or without sample bias. The results show that measuring without bias and with the flood gun, is the optimum measurement condition resulting in a stabilized surface potential and repeatable measurements. This is shown in Figure 12a, when a sample of 1 cm of the model resist is measured for 2 hours and the UPS spectra recorded every ~2 minutes.

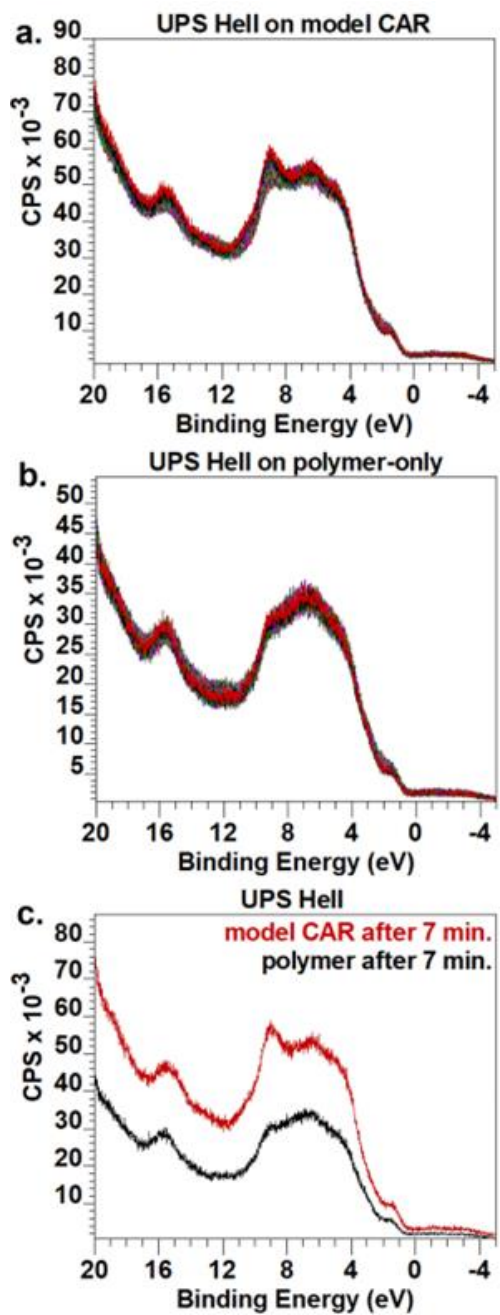

Figure 12: UPS HeII: a) repeated short measurements on model CAR, recorded for a total duration of 2 hrs; b) repeated short measurements on polymer-only, recorded for a total duration of 2 hrs; c) a comparison between the recorded spectra on model CAR and polymer, after a total measurement time of 7 min for each sample.

It is clear that all the 56 recorded spectra overlap nicely in the whole binding energy region except in the region between 7 to 12 eV. It is clearly visible that the sharp peak at around 9 eV in the first spectrum degrades gradually during 2 hrs measurement, which might be an indication of activation of the resist and photochemical changes during 2 hours exposure to HeII photons.

This is confirmed with the result of 2 hours measurement on polymer-only sample, with the same protocol, where no significant changes are visible between all the recorded spectra. Also, the shape of these spectra is quite similar to recorded spectrum for the model CAR, after 2 hrs measurement. This observation in the model CAR, might imply the presence of only polymer and absence of PAG and quencher in the system, after 2 hrs exposure to UV photons. However, the experimental data is not enough to understand the details of this photochemistry. On the other hand, it seems that eliminating the degradation of the sample under UPS measurements in our setup would be possible by short measurements of about 7 minutes, while having an acceptable SNR.

To understand the details of this chemical modification in the model resist, simulations were carried out to predict the DOS of the system before and after chemical changes, and the results were compared to the experimental spectra as shown in Figure 13. To make a valid comparison, it was needed to align the simulated spectra with the experimental one. However, before this alignment, a correction is required on the binding energy position of the experimental spectrum to make sure that the spectra are measured with respect to the vacuum level and that the zero binding energy is compatible with the experimental Femi level. To this end, first a sample of polymer-only was measured with XPS to obtain the valence band spectrum as well as C1s. The reason for using polymer-only sample instead of model CAR for this XPS measurement, is the significant slower degradation of the polymer-only rather than the model CAR. This is important especially due to low ionization cross section of XPS in the valence band region, which should be compensated by long measurement in order to obtain an accepted SNR. Considering that more than 80 mol% of the ESCAP material is polymer, this approach seems reasonable.

Then the valence band spectrum was calibrated with the customary method of referencing by C1s spectrum. Later the experimental UPS (HeII) spectra of the model CAR, were shifted for 2.4 eV towards higher binding energies, to align with the corrected XPS valence band spectrum, by using similar features in both spectra. The UPS spectra with calibrated binding energy, after 2 minutes and 2 hours UPS measurement are illustrated in Figure 13, with the sharp peak positioned at around 11 eV.

The simulated UPS spectra (based on DOS calculations) were also shifted to align with the experimental spectra as shown Figure 13.

After binding energy correction, comparison between similar features in experimental and simulated spectra helps to interpret the photochemistry. The simulated model considers different pathways for fragmentation of the model CAR. It appears that the fragmentation of both PAG and PBMA component of the co-polymer (blue curve) represents best the experimental spectra. This assumption is also in line with the traditional activation pathway of the model CAR under UV exposure which includes the PAG activation, acid generation and diffusion in the system, which will trigger the fragmentation of PBMA through the so called deprotection process.

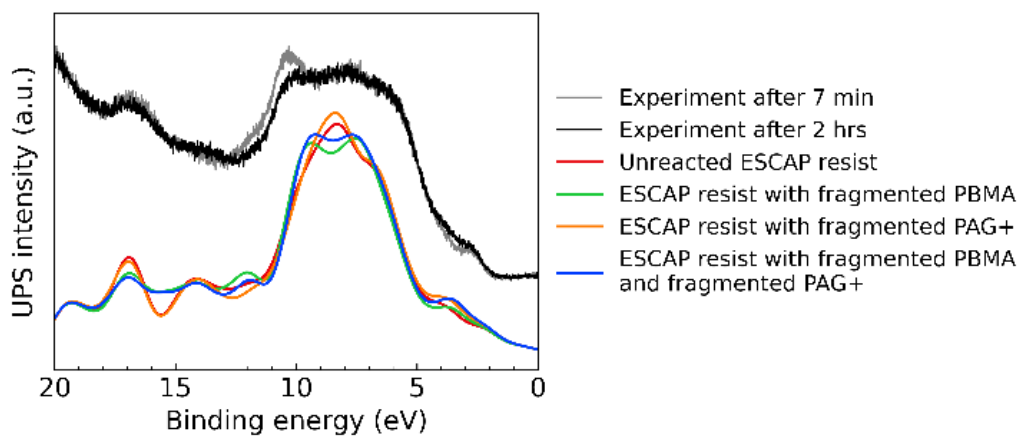

Figure 13: Simulated UPS spectra calculated based on theoretical DOS, vs the experimental spectra for the model CAR. The simulated spectra were shifted to align the main peaks to the experimental spectra.

# IV. Conclusion

In photolithography, a pattern is transferred into a solid material using a photosensitive material (photoresist) that converts the light signal into a physical barrier (i.e. for further etch). The basic working principle of the photoresists is the solubility change in the parts that are exposed to light, so that in a subsequent development step, a solvent selectively removes either the exposed (positive tone development) or the unexposed part (negative tone development). Several types of photoresists have been developed for different wavelengths of light which employ different photochemistry for the activation of the solubility switch mechanism.

A commonly used type of photoresist for different ranges of wavelengths is the chemically amplified resist (CAR) which relies on chemical amplification, using a catalyst. A CAR is composed of a polymer backbone, a photoacid generator (PAG) and a quencher. In CAR, PAG molecules are activated during light exposure and generate acid.

Then during a subsequent post-exposure bake (PEB) the generated acid catalyzes a deprotection reaction on the side chains of the polymer matrix which leads to the solubility change of the polymer. The advantage of catalyst used in CARs, is the high deprotection efficiency, i.e., few PAG molecules doped into the resist matrix, can activate thousands of deprotection reactions.

In this work, we measured a model CAR with XPS and UPS techniques and evaluated the sample modification during the measurement.

We show that photoemission analysis of photoresist should be treated with extreme attention to sample degradation induced by the measurements, especially in the presence of photon induced catalyst species.

The damage during XPS measurement can for instance be visualized under an optical microscope after a development procedure. AFM topography and XPS elemental scanning were used for further investigations about the nature of this damage. AFM confirmed the change in the thickness of the CAR layer in the exposed area to X-ray. The highest thickness reduction was observed at the edge of modified area and indicates full development and removal of the CAR in this area. This is confirmed by detection of high concentration of Si substrate in the light ring area with XPS elemental mapping. AFM measurements also show a random variation in the thickness of the photoresist in the centre of the damage spot, with a maximum of 20 nm thickness reduction. In addition, XPS elemental mapping in the center of the damage area, detects no silicon but high concentration of carbon which is most probably from the polymer backbone and originates from non-developed photoresist or cross-linked polymer from overexposure of the CAR to high energy X-ray photons.

The results of theoretical simulations also support the hypothesis of PAG dissociation during the XPS measurements and confirm the degradation of CAR during XPS measurements.

An XPS measurement protocol was proposed to alleviate/limit the measurement induced modification by spreading the X-ray photon beam over a larger measurement area than what is used in standard XPS. This protocol is technically applied by rastering the beam over a larger measurement spot. It is shown that it allows measurement up to 45 min with no indication of modification.

Further, during UPS measurements on the same model ESCAP, sample modification was detected as degradation in a peak at ~11 eV, after 2 hours measurement. A simulated model of the UPS spectrum on the same system suggests that this degradation could be related to activation of PAG, followed by fragmentation of PBMA component of the co-polymer. UPS can however be performed with sufficient SNR and without noticeable degradation for short measurements (less than or equal to 7 minutes) in our UPS setup.

## ACKNOWLEDGMENTS

We would like to express our sincere gratitude to Paul van der Heide and John Petersen for their kind support.

K.M.D and D. P. S. acknowledge funding from the European Union's Horizon 2020 research and innovation programme under the Marie Sklodowska-Curie grant agreement No.'s 101031245 (K.M.D.) and 101032241 (D.P.S.).

# DATA AVAILABILITY

The data that support the findings of this study are available from the corresponding author upon reasonable request.